\documentclass[12pt]{elsarticle}
\usepackage{ifpdf,amssymb,amsmath,epsfig,bm,pifont}
\usepackage{graphics}
\usepackage{graphicx}
\usepackage{tabularx}
\usepackage{multirow}
\usepackage{dcolumn}
  \graphicspath{{./figs/}}
   \pdfoutput=1
\usepackage{color}

\def\be{\begin{equation}}
\def\ee{\end{equation}}
\def\bea{\begin{eqnarray}}
\def\eea{\end{eqnarray}}
\def\bear{\begin{array}}
\def\eear{\end{array}}
\def\bes{\begin{subequations}}
\def\ees{\end{subequations}}
\newcommand{\MSbar}{\overline{\rm MS}}  
\newcommand{\m}{{\overline m}}
\newcommand{\A}{{\mathcal{A}}}
\newcommand{\tA}{{\widetilde {\mathcal{A}}}}

\newcommand{\tk}{{\widetilde k}}

\renewcommand{\thefootnote}{\fnsymbol{footnote}}

\biboptions{compress}

\begin{document}
\begin{frontmatter}

\title{\vskip-3cm{\baselineskip14pt
\centerline{\normalsize\hfill FTUV-15-0623.3799, IFIC/15-40, USM-TH-336}
}\vskip.7cm
\texttt{anQCD}: \texttt{Fortran}  
programs for couplings at complex momenta in various analytic QCD models
\vskip.3cm
}
 \author[ificUV]{C\'esar Ayala}
\ead{c.ayala86@gmail.com}

 \author[utfsm]{Gorazd Cveti\v{c}}
\ead{gorazd.cvetic@usm.cl}

\address[ificUV]{Department of Theoretical Physics and IFIC,\\ 
University of Valencia and CSIC, E-46100, Valencia, Spain}
\address[utfsm]{Department of Physics, Universidad T\'ecnica Federico Santa Mar\'ia,\\
                Casilla 110-V, Valpara\'iso, Chile\\}

\date{}

\begin{abstract}
\noindent
We provide three \texttt{Fortran} programs which 
evaluate the QCD analytic (holomorphic) couplings $\mathcal{A}_{\nu}(Q^2)$
for complex or real squared momenta $Q^2$.
These couplings are holomorphic analogs of the powers
$a(Q^2)^{\nu}$ of the underlying perturbative QCD (pQCD) coupling
$a(Q^2) \equiv \alpha_s(Q^2)/\pi$,
in three analytic QCD models (anQCD):
Fractional Analytic Perturbation Theory (FAPT), 
Two-delta analytic QCD (2$\delta$anQCD), 
and Massive Perturbation Theory (MPT).
The index $\nu$ can be noninteger.
The provided programs do basically the same job as the \texttt{Mathematica} package
\texttt{anQCD.m} published by us previously,
Ref.~\cite{CPC1_CAGC},
but are now written in \texttt{Fortran}.

\vspace{.2cm}


\end{abstract}

\end{frontmatter}

\thispagestyle{empty}
 \newpage
  \setcounter{page}{1}

\renewcommand{\thefootnote}{\arabic{footnote}}
 \setcounter{footnote}{0}

\section*{Program Summary}
\begin{itemize}

\item[]\textit{Program titles:}
  \texttt{AFAPText.for, A2danQCDext.for, AMPText.for}

\item[]\textit{The main programs (}\texttt{AFAPText.for, A2danQCDext.for, AMPText.for}\textit{) 
and the tar-gzipped file containing all three programs (}\texttt{AanQCDextFOR.tar.gz}\textit{), available from the web page:}\\  \texttt{gcvetic.usm.cl}

\item[]\textit{Computer:}
  Any work-station or PC where \texttt{Fortran 95/2003/2008} (gfortran) is running

\item[]\textit{Operating system in which the program has been
    tested:}
  Operating system Linux (Ubuntu and Scientific Linux), Windows (in all cases
using gfortran)

\item[]\textit{No. of bytes in distributed programs:}\\
  103 kB (\texttt{AFAPText.for}), 107 kB (\texttt{A2danQCDext.for}), 236 kB  (\texttt{AMPText.for}); 99 kB (\texttt{AanQCDextFOR.tar.gz})

\item[]\textit{Distribution format:}
  tar.gz
\item[]\textit{Keywords:}
  Analytic (holomorphic) QCD coupling,
  Fractional Analytic Perturbation Theory,
  Two-delta analytic QCD model,
  Massive Perturbation Theory,
  Perturbative QCD,
  Renormalization group evolution.

\item[]\textit{Nature of problem:}
 Calculation of values of the running analytic couplings 
 $\A_{\nu}(Q^2;N_f)$ for general complex squared momenta $Q^2 \equiv -q^2$,
 in three analytic QCD models, where  $\A_{\nu}(Q^2;N_f)$
 is the analytic (holomorphic) analog of the power 
 $(\alpha_s(Q^2;N_f)/\pi)^{\nu}$. Here, $\A_{\nu}(Q^2;N_f)$ is a holomorphic
 function in the $Q^2$ complex plane, with the exception of the negative
 semiaxis $(-\infty,-M^2_{\rm thr})$, reflecting the analiticity properties
 of the spacelike renormalization invariant quantities ${\cal D}(Q^2)$ in
 QCD. In contrast, the perturbative QCD power 
 $(\alpha_s(Q^2;N_f)/\pi)^{\nu}$ has singularities even outside the
 negative semiaxis (Landau ghosts).  The three considered models are:
 Analytic Perturbation theory (APT); Two-delta 
 analytic QCD (2$\delta$anQCD); Massive Perturbation Theory (MPT).
 We refer to Ref.~\cite{CPC1_CAGC} for more details and literature.

\item[]\textit{Solution method:}
  The \texttt{Fortran} programs for FAPT and 2$\delta$anQCD models contain 
  routines and functions needed to perform two-dimensional numerical 
  integrations involving the spectral function, in order to evaluate
  $\A_{\nu}(Q^2)$ couplings. In MPT model, one-dimensional numerical integration
  involving $\A_{1}(Q^2)$ is sufficient to evaluate any $\A_{\nu}(Q^2)$ coupling.

\item[]\textit{Restrictions:}
  For unphysical choices of the input parameters
  the results are meaningless. When $Q^2$ is close to the cut region of
  the couplings ($Q^2$ real negative), the calculations can take more time
  and can have less precision.

\item[]\textit{Running time:}
  For evaluation of a set of about 10 related couplings, the times vary 
  in the range $t \sim 10^1$-$10^2$ s. 
  MPT requires less time, $t \sim 1$-$10^1$ s.

\end{itemize}

\newpage


\section{General remarks}
\label{sec:gr}

The included \texttt{Fortran} programs evaluate the holomorphic couplings
$\A_{\nu}(Q^2)$ in the dispersive approach to QCD, where we consider 
three analytic versions: Fractional Analytic Perturbation Theory (FAPT) 
\cite{FAPT} (and references therein), 2$\delta$ analytic QCD (2$\delta$anQCD) 
\cite{2danQCD} (cf.~also \cite{CPC1_CAGC,anOPE}),
and Massive Perturbation Theory (MPT) \cite{MPT}
(cf.~also \cite{Simonov,BadKuz}).
These programs are constructed for complex squared momenta
$Q^2 \equiv - q^2$, in a similar way as the previously published
Mathematica package anQCD.m, Ref.~\cite{CPC1_CAGC} (also on the web
page: gcvetic.usm.cl).\footnote{
The analogous \texttt{Fortran} programs, but only for real values of the squared
momenta $Q^2$, were mentioned in Ref.~\cite{ACAT_CAGC} 
(and were made available on the web page: gcvetic.usm.cl). 
The \texttt{Fortran} program for the couplings
$A_n(Q^2)$ for integer index $n$ and real $Q^2$ in APT \cite{APT} 
and massive APT (MAPT) \cite{MAPT} was 
provided in Ref.~\cite{NestFor},
based on the corresponding program in Maple \cite{NestMap}. 
The Mathematica program for evaluation of the
general power analogs $\A_{\nu}(Q^2)$ in FAPT was provided
in Ref.~\cite{BK}.} 
For additional details on the three analytic QCD models with a view
to implementing them numerically in Mathematica, we refer to
Ref.~\cite{CPC1_CAGC}.

A complication appears in \texttt{Fortran}, though,
because in the dispersion integrals for FAPT and $2\delta$anQCD
couplings the polylogarithm function ${\rm Li}_{\nu}(z)$
for complex $z$ appears and it has not been implemented in
\texttt{Fortran} yet (in \texttt{Mathematica} 
it is implemented as ${\rm PolyLog}[-\nu,z]$).
In the attached \texttt{Fortran} programs 
this function is calculated as the following integral 
\cite{Kotikov:2010bm}:
\be
{\rm Li}_{-n-\delta}(z) = \left( \frac{d}{d \ln z} \right)^{n+1}
\left[ \frac{z}{\Gamma(1 - \delta) } \int_0^1 \frac{d \xi}{1 - z \xi}
\ln^{-\delta} \left( \frac{1}{\xi} \right) \right] \quad
(n=-1, 0, 1, \ldots; \ 0 < \delta < 1) \ ,
\label{Li_nugen}
\ee
where $\nu=n+\delta$. For better stability, the numerical integration of
this integral over $\xi$ is implemented in the complex plane along a ray
in the first quadrant. This then results in two-dimensional
integration for the evaluation of FAPT and $2\delta$anQCD couplings
$\A_{\nu}(Q^2)$. This integration is performed numerically with
vegas routine \cite{vegas}. In this approach, overflow and/or underflow
problems appear at the edges of the unit square of integration, 
which are dealt with carefully in the programs. 
In the evaluation of couplings $\A_{\nu}(Q^2)$ in MPT model, one-dimensional 
integrations are needed because the integrand involves
$\A_{1}(Q^2/\xi)$ and no functions ${\rm Li}_{\nu}(z)$.

The \texttt{Fortran} programs are self-contained, i.e., no additional
packages are needed. The needed explanations and instructions on
the input parameters, compiling commands, and the output form, are all
given at the beginning of each \texttt{Fortran} program.
The calculations are more time-consuming in FAPT when the
complex squared momenta $Q^2$ are close to the cut negative semiaxis.
In all models (FAPT, $2\delta$anQCD, MPT), the number of active quark flavors
$N_f$ is a fixed input integer. 

We wish to point out that this program, in \texttt{Mathematica} form
\cite{CPC1_CAGC}, was already applied without problems
to the evaluation of the nonsinglet structure function $F_2$ in
deep inelastic scattering in FAPT model \cite{Ayala:2015epa}.

\section{Practical aspects of the program}
\label{sec:pap}

\subsection{Input parameters}
\label{subs:inp}

The programs can get compiled with the simple ``gfortran'' command.
For example, the program A2danQCDext.for can get compiled by
writing in the directory where the program is stored: 
``gfortran -o A2dan A2danQCDext.for'' and then executed with the
command ``./A2dan''.
Before compiling, however, the input parameters should be typed into
the program, at (two) places which start with the string ``INPUT''.
In all three programs the following input parameters need to be specified: 
\begin{enumerate}
\item
$N_f$ (``Nf''), the number of active quarks; 
\item
$N_{\rm in}$ (``Nin'') and $\delta_{\rm in}$ (``delin'') indices, 
where $\nu=N_{\rm in}+\delta_{\rm in}$
is the index of the coupling $\A_{\nu}$;
where $N_{\rm in}=0,1,2,3,4$; and  $0 \leq \delta_{\rm in} \leq 1$.
\item
$|Q^2|$ (``AbsQ2'', in ${\rm GeV}^2$) and $\phi$ (``ArgQ2'', in radians),
where $Q^2 = |Q^2| \exp(i \phi)$ is the (complex) squared momentum
($Q^2 \equiv - q^2$).
\end{enumerate}
In addition, in the programs AFAPText.for and AMPText.for, the
scale ${\overline \Lambda}_{N_f}$ 
(``gL2MS'', in GeV) of the underlying pQCD coupling needs to be specified.

Further, in AMPText.for, the MPT squared effective mass of the gluon 
$m^2_{\rm gl}$ (``gM2'', in ${\rm GeV}^2$) needs to be specified. 
This mass is of the order of magnitude
$m^2_{\rm gl} \sim 1 \ {\rm GeV}^2$ \cite{MPT,Simonov,BadKuz},
and is assumed in our program AMPText.for to be constant at all
momenta (and thus independent of $Q^2$ and of $N_f$).
Instead of the input scale  ${\overline \Lambda}_{N_f}$ in MPT,
which basically determines the strength of the MPT coupling
$\A_1^{\rm (MPT)}(Q^2) = \alpha_s(Q^2+m^2_{\rm gl};\MSbar)/\pi$,
one might prefer to use the value of $\pi \A_1(M_Z^2) =
\alpha_s(M_Z^2+m^2_{\rm gl};\MSbar)$. This then determines the value
of the scale ${\overline \Lambda}_{N_f=5}$.  The values
of other scales ${\overline {\Lambda}}_{N_f}$ (for $N_f=4,3,6$) can
then be obtained by applying the (3-loop) quark threshold relations \cite{thr4l}
applied within the (analytic) MPT model
\begin{eqnarray}
\label{msbMPT}
\A_1^\prime &=&\A_1- \A_2 \frac{\ell_h}{6}
+ \A_3 \left(\frac{\ell_h^2}{36}-\frac{19}{24}\ell_h+ {\widetilde c}_2\right)
\nonumber\\
&& 
+ \A_4 \left[ -\frac{\ell_h^3}{216}
-  \frac{131}{576}\ell_h^2+\frac{\ell_h}{1728} \left( -6793+281 (N_f-1) \right)
+ {\widetilde c}_3 \right].
\end{eqnarray}
Here, $\A_1^\prime \equiv \A_1^{\rm (MPT)}(\mu^2_{N_f};N_f-1)$ and
$\A_n \equiv \A_n^{\rm (MPT)}(\mu^2_{N_f};N_f)$,
and $\ell_h=\ln[\mu_{N_f}^2/{\m}_q^2] = \ln \kappa$ where ${\m}_q={\m}_q({\m}_q)$
is the $\MSbar$ mass of the corresponding quark entering at the
threshold squared momentum $\mu_{N_f}^2 = \kappa {\m}_q^2$ ($\kappa \sim 1$).
We recall that in MPT (and FAPT) we use for the underlying
pQCD coupling the 4-loop $\MSbar$ running coupling.
In Table \ref{T1} we present the values of the scales
${\overline {\Lambda}}_{N_f}$ corresponding to various values
of $m^2_{\rm gl}$ and $\pi \A_1(M_Z^2)$ in MPT.
A similar Table for the values of the scales 
${\overline {\Lambda}}_{N_f}$ was given for FAPT
in Ref.~\cite{CPC1_CAGC} (Table 1 there). 
\begin{table}
\caption{The scales ${\overline {\Lambda}}_{N_f}$, written in MeV,
in MPT analytic model, for various values of $m^2_{\rm gl}$
and $\pi \A_1^{\rm (MPT)} = \alpha_s(M_Z^2+m^2_{\rm gl};\MSbar)$.
The threshold parameter $\kappa=2$ was taken; in parentheses, the results
with the threshold parameter $\kappa=1$ are given.}
\label{T1}
\begin{center}
\centering
\begin{tabular}{|c c|c c c c}
\hline
$m^2_{\rm gl}$ $[{\rm GeV}^2]$ & $\pi \A_1^{\rm (MPT)}(M_Z^2)$ &
${\overline {\Lambda}}_{6}$ & ${\overline {\Lambda}}_{5}$ &
${\overline {\Lambda}}_{4}$ & ${\overline {\Lambda}}_{3}$ 
\\
\hline
0.5 & 0.118 & 88.3 (88.4) & 208.6 (208.6) & 291.5 (291.0) & 339.8 (338.4)
\\
1.0 & 0.118 & 88.3 (88.4) & 208.6 (208.6) & 291.8 (291.3) & 343.1 (342.1)
\\
1.5 & 0.118 & 88.3 (88.4) & 208.7 (208.7) & 292.1 (291.7) & 346.5 (345.4)
\\
\hline
0.5 & 0.120 & 99.8 (99.8) & 232.9 (232.9) & 321.8 (321.2) & 371.1 (369.5)
\\
1.0 & 0.120 & 99.8 (99.8) & 233.0 (233.0) & 322.1 (321.6) & 374.8 (373.6)
\\
1.5 & 0.120 & 99.8 (99.9) & 233.0 (233.0) & 322.5 (321.9) & 378.5 (377.2)
\\
\hline
0.5 & 0.122 & 112.3 (112.4) & 259.1 (259.1) & 354.0 (353.3) & 404.2 (402.1)
\\
1.0 & 0.122 & 112.3 (112.4) & 259.1 (259.1) & 354.4 (353.8) & 408.0 (406.7)
\\
1.5 & 0.122 & 112.3 (112.4) & 259.1 (259.1) & 354.8 (354.2) & 412.1 (410.6)
\\
\hline
\end{tabular}
\end{center}
\end{table}

In A2danQCD.for program for 2$\delta$anQCD model, 
the scheme parameter $c_2$ ($=\beta_2/\beta_0$)
was set equal to the central preferred value $c_2=-4.9$
(cf.~Table 2 of Ref.~\cite{CPC1_CAGC}).

\subsection{Output}
\label{subs:outp}

After the execution of the program, the results of each program are written
in the output file \texttt{AFAPText.dat} (or: \texttt{A2danQCDext.dat}, \texttt{AMPText.dat}). 
If the input index is $\nu=N+\delta$
($N=0,1,2,3$ or $4$; and $0 \leq \delta \leq 1$), 
the output will consist of the following couplings:
\bes
\bea
&& \tA_{N+\delta}(Q^2), \; \tA_{N+1+\delta}(Q^2), \; \ldots , \; \tA_{4+\delta}(Q^2),
\label{tAs}
\\
&& \A_{N+\delta}(Q^2) ({\rm N}^{N}{\rm LO}), \; 
\A_{N+\delta}(Q^2) ({\rm N}^{N+1}{\rm LO}), \; \ldots ,
\A_{N+\delta}(Q^2) ({\rm N}^{4}{\rm LO}).
 \label{As}
\eea
\ees
Here we recall that $\tA_{\nu}(Q^2)$ are the logarithmic derivatives
analytically extended to noninteger index $n \mapsto \nu$, where for integer
$n$ these derivatives are
\be
{\tA}_{n}(Q^2) \equiv \frac{(-1)^{n-1}}{\beta_0^{n-1} (n-1)!}
\left( \frac{ \partial}{\partial \ln Q^2} \right)^{n-1}
\A_1(Q^2) \ , \qquad (n=1,2,\ldots) \ .
\label{tAn}
\ee
The couplings $\tA_{\nu}(Q^2)$ are analytic (holomorphic) in $Q^2$, and
perturbatively we have $\tA_{\nu} \sim (\alpha_s(Q^2)/\pi)^{\nu}$.
Further, the couplings $\A_{\nu}(Q^2)$ are the analytic (holomorphic) 
analogs of the powers $(\alpha_s(Q^2)/\pi)^{\nu}$, and they are
linear combination of the couplings $\tA_{\nu+K}(Q^2)$ ($K=0,1,\ldots$). 
For example, $\A_{N+\delta}(Q^2)$ at ${\rm N}^{N+M}{\rm LO}$ precision 
($N$ is integer; $0 \leq \delta \leq 1$) is the following linear combination:
\bea
\A_{N+\delta}(Q^2) & = & \tA_{N+\delta}(Q^2) + \sum_{m=1}^{M} \tk_m(N+\delta)
\tA_{N+m+\delta}(Q^2).
\label{ANdel}
\eea
The expressions for the general coefficients $\tk_m(N+\delta)$ were obtained in
Ref.~\cite{GCAK}. These coefficients involve  the Euler $\Psi$ function
$\Psi(\nu) = (d/d \nu) \Gamma(\nu)$ and its derivatives $\Psi^{(m)}(\nu)$.
In \texttt{Fortran}, $\Psi(\nu)$ and $\Psi^{(m)}(\nu)$ are calculated by the
routine \cite{PsiFort} which is used in our \texttt{Fortran} programs.

\vspace*{+7mm}
\textbf{Acknowledgments}\vspace*{+1mm}
This work was supported by FONDECYT (Chile) Grant No. 1130599 and DGIP 
(UTFSM) internal project USM No. 11.15.41 (G.C), 
Spanish Government and ERDF funds from the EU Commission [FPA2014-53631-C2-1-P, No. CSD2007-00042(Consolider Project CPAN)] and by CONICYT Fellowship ``Becas Chile'' Grant No.74150052 (C.A).

\end{document}